%% file: main_arxiv.tex
\documentclass[letterpaper, 10 pt, conference]{ieeeconf}  % Comment this line out
\IEEEoverridecommandlockouts                              % This command is only
\input{PackagesCommands}
\usepackage{multicol,lipsum}
\usepackage[export]{adjustbox}
\usepackage[acronym]{glossaries}
\newacronym{nn}{NN}{Neural Network}
\newacronym{gd}{GD}{Gradient Descent}
\newacronym{fs}{FS}{Fsolve}
\newacronym{rsm}{RSM}{Random Search Method}
\newacronym{trd}{TRD}{Trust-Region-Dogleg}

\tikzstyle{rectRound} = [rectangle, rounded corners, text centered, draw=black, minimum width=2em]
\tikzstyle{rect} = [rectangle, text centered, draw=black, minimum width=2em]
\tikzstyle{clear} = [draw=none, text centered, minimum width=2em]
\tikzstyle{circ} = [circle, text centered, draw=black, minimum width = 2em]
\tikzstyle{circClear} = [circle, text centered, minimum width = 2em]
\tikzstyle{arrow} = [thick, -latex]
\makeatletter
\DeclareRobustCommand{\rvdots}{%
  \vbox{
    \baselineskip4\p@\lineskiplimit\z@
    \kern-\p@
    \hbox{.}\hbox{.}\hbox{.}
  }}
\makeatother

\usepackage[style=ieee ,sorting=none,maxbibnames=99,giveninits, doi=false, backend=biber]{biblatex}
\title{\LARGE \bf
Longitudinal Flight Dynamics Control Based on Feedback Linearization and Normal Canonical Form
}

\title{\LARGE \bf
MIMO Input-Output Linearization with Applications for Longitudinal Flight Dynamics
}

\title{\LARGE \bf
A
Tutorial on Neural Networks and Gradient-free Training
}

\author{Turibius Rozario, Arjun Trivedi, Ankit Goel% \thanks{The University of Maryland, Baltimore County}% <-this % stops a space
\thanks{Turibius Rozario is an undergraduate student and a Meyerhoff Scholar in the Department of Mechanical Engineering, University of Maryland, Baltimore County, 1000 Hilltop Circle, Baltimore, MD 21250. {\tt \small s175@umbc.edu}}%
\thanks{Arjun Trivedi graduated from the Department of Mechanical Engineering, University of Maryland, Baltimore County, 1000 Hilltop Circle, Baltimore, MD 21250. {\tt\small atrived2@umbc.edu}}%
\thanks{Ankit Goel is an Assistant Professor in the Department of Mechanical Engineering, University of Maryland, Baltimore County,1000 Hilltop Circle, Baltimore, MD 21250. {\tt\small ankgoel@umbc.edu }}%
}
\usepackage{graphicx}      % include this line if your document contains figures
\usepackage{textcomp}
\usepackage{calc}
\usepackage{mathtools}
\usepackage{xparse}
\usepackage{amssymb}
\usepackage{amsmath}
\usepackage{cases}
\usepackage{mathtools}
\usepackage{cuted}

\usepackage{breakurl}

\begin{document}

\maketitle
\thispagestyle{empty}
\pagestyle{empty}

%%%%%%%%%%%%%%%%%%%%%%%%%%%%%%%%%%%%%%%%%%%%%%%%%%%%%%%%%%%%%%%%%%%%%%%%%%%%%%%%
\begin{abstract}
This paper presents a compact, matrix-based representation of neural networks in a self-contained tutorial fashion. 
% 
% Specifically, we develop neural networks as a composition of several vector-valued functions. 
% 
Although neural networks are well-understood pictorially in terms of interconnected neurons, neural networks are mathematical nonlinear functions constructed by composing several vector-valued functions. 
Using basic results from linear algebra, we represent a neural network as an alternating sequence of linear maps and scalar nonlinear functions, also known as activation functions. 
% , which are parameterized by matrix multiplications, and nonlinear maps. 
% 
The training of neural networks requires the minimization of a cost function, which in turn requires the computation of a gradient. 
Using basic multivariable calculus results, the cost gradient is also shown to be a function composed of a sequence of linear maps and nonlinear functions.
% also known as backpropagation.   
% 
In addition to the analytical gradient computation, we consider two gradient-free training methods and compare the three training methods in terms of convergence rate and prediction accuracy.  
% Three gradient-free training schemes are implemented and the results of training are compared with the gradient-based training. 
% Finally, a novel data-drive, gradient free training algorithm is presented and compared. 

\end{abstract}

%%%%%%%%%%%%%%%%%%%%%%%%%%%%%%%%%%%%%%%%%%%%%%%%%%%%%%%%%%%%%%%%%%%%%%%%%%%%%%%%
\section{INTRODUCTION}

Neural networks, modeled and named after millions of interconnected neurons in our brains, were first introduced in 1940s. 
Over the last decade, neural networks have found tremendous success in almost every domain of science and engineering. 
Neural networks have enabled natural language processing, speech recognition, image search, spam classification, and autonomous navigation to name just a few \cite{shinde2018review, bagnell2010learning,tullu2021machine}. 

% auto-encoders, prediction, and computer vision;
% as everyday examples, Microsoft's speech recognition, Google's image search service, and
% Facebook's spam message clean up are made using neural networks \cite{shinde2018review}.
% In the physical realm, machine learning can be used to train neural networks for 
% autonomous navigation for unmanned land \cite{bagnell2010learning} and aerial \cite{tullu2021machine} vehicles.

The key technology that has accelerated the success rate of neural networks is the precipitous drop in the cost of computation.
Specifically, the increase in computational speed and the simultaneous increase in the efficient use of memory and storage has removed the barriers that hindered the progress of neural networks for almost five decades since their inception in the 1940s.

% Neural networks have been used in the financial prediction application \cite{xing2018natural}

Although neural networks have been interpreted in numerous ways using anatomical concepts, modern neural networks are an extremely large composition of mathematical functions, often parameterized by millions of parameters or \textit{gains}. 
The input to a neural network is often a mathematical vector, therefore,  physical inputs such as images or sound clips are converted into mathematical vectors before being passed to the neural network. 
The output of a neural network similarly is a mathematical vector. 
Depending on the application, the neural network's output can be assigned a physical meaning such as an object or the probability of an event. 
Neural networks are \textit{trained} by minimizing a cost function constructed using the prediction error.

Since neural networks are nonlinearly parameterized by their gains, the resulting optimization problem does not possess a closed-form analytical solution. 
Numerical techniques based on the gradient of the cost function are therefore used to train neural networks, which is computationally the most expensive part of the training process. 
However, the gradient of the cost function can be computed using an analytical closed-form solution since modern neural networks are constructed using well-behaved functions.
The gradient computation requires evaluation of the neurons in the neural network starting from the last layer and proceeding backward, a process often called \textit{backpropagation}.

% Using large training datasets, neural networks are \textit{trained} by minimizing a cost function 

To reduce the computational cost of training neural networks,  
several techniques have been developed over the last two decades.
% 
% For computer vision, pooling and convolution improve both neural network accuracy and rate of learning.
Optimizers such as stochastic gradient descent with momentum, RMSprop, and Adam have been shown to improve the convergence rate of the neural network gains by adjusting the learning rate during training \cite{goodfellow2016deep}.
In addition to algorithmic improvements, neural network training has also benefited from hardware improvements such as the use of GPUs to conduct neural network training computations \cite{steinkraus2005using}.

This paper aims to present a compact, matrix-based representation of neural networks in a tutorial fashion. 
Specifically, we show that a neural network is constructed by composing linear maps with nonlinear scalar functions.
Due to the flexibility in the design of a neural network, neural network gains are not vectors, but a set of matrices. 
Furthermore, we show that the gradient of the cost, which requires the gradient of individual layer outputs with respect to the neural layer gains, is also constructed by composing linear maps with nonlinear scalar functions.
In addition to the gradient-based training, we also present two gradient-free training methods. 
The first method is based on the root-finding problem. By recognizing the neural network output prediction as a system of nonlinear equations, we apply standard root-finding techniques to train the neural network.  
The second method is motivated by the simulated annealing technique, often used in nonlinear system identification. 
Instead of computing gradient to determine the direction in which to update the neural network gains, we generate an ensemble of gains normally distributed on a hypersphere around the latest estimate of the gains. 
The process is repeated by computing the cost for each ensemble member and choosing the ensemble member with the minimum cost until satisfactory prediction accuracy is obtained.

Although many extensions of the basic neural network have been developed for a variety of applications, in this tutorial paper, we solely focus on simple neural networks, which are usually the building blocks of complex networks such as recurrent neural networks, convolutional neural networks, and generative adversarial networks.

% The paper is intended as a tutorial for neural networks. 

% Applications/importance

% Past/recent work

% Contributions

The paper is organized as follows. 
Section \ref{sec:neural_networks} presents the mathematical form of the neural networks,
Section \ref{sec:NN_training} presents a gradient-based and two gradient-free methods to train neural networks, 
and 
Section \ref{sec:examples} presents three examples that compare the convergence rate of the three training methods. 

%% Sections/NN.tex copy pasted below
\section{Neural Networks}
\label{sec:neural_networks}

This section presents a compact matrix-based representation of neural networks.
% A neural network is composed of several \textit{neural layers}, and each neural layer is composed of several \textit{neurons}.
% 
Specifically, in this section, we write neurons and neural layers as mathematical functions and construct neural networks as composition of several neural layers. 

A \textit{neuron} is the most basic component of a neural network. 
Mathematically, the input to a neuron is a real-valued vector, and its output is a scalar. 

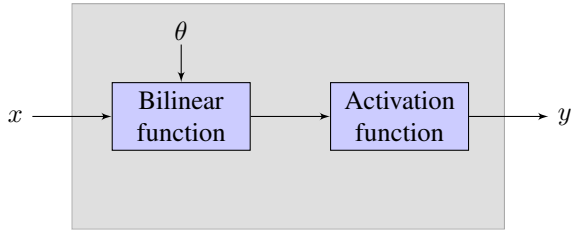
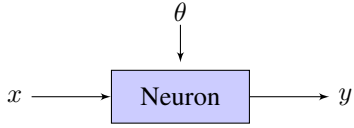
\begin{figure}
    \centering
    \subfloat[Anatomy of a neuron.]
    {
    \centering
    % \resizebox{\columnwidth}{!}
    {
    \begin{tikzpicture}[auto, node distance=2cm,>=latex',text centered]
    
        \draw [draw=black, fill = gray, opacity=0.25] (-2.25,-1.5) rectangle (3.5,1.5);
        \node at (-3,0) (input) {$x$};
        \node [smallblock, right = 3 em of input,text width=1.6cm] (bilinear) {Bilinear function};
        % , minimum height=3em, text width=1.6cm
        
        \node [smallblock, right = 3 em of bilinear,text width=1.6cm] (activation) {Activation function};
        % \node at {(Plant)+(3,0)}  (output) {$y$};
        \node[right = 3 em of activation] (output) {$y$};
        
        \draw[->] (input) -- (bilinear);
        \draw[->] (bilinear) -- (activation);
        \draw[->] (activation) -- (output);
        % \draw[->] (sum) -- node[xshift = 0em, yshift = .2em]{$e$} (Controller);
        % \draw[->] (Controller)-- node[xshift = 0em, yshift = .2em]{$u$} (Plant);
        % \draw[->] (Plant) -- (output);
        % \draw[->] (Plant.east) -- +(1,0) -- +(+1,-1.5)
        %             -| (sum.south);
        \draw[->] (bilinear.north)+(0,0.5) node[xshift = 0em, yshift = +0.5em]{$\theta$} -- (bilinear.north);

    \end{tikzpicture}
    }
    }
    \\ 
    \subfloat[Simplified representation of a neuron.]
    {
    \centering
    % \resizebox{\columnwidth}{!}
    {
    \begin{tikzpicture}[auto, node distance=2cm,>=latex',text centered]
        \node at (-3,0) (input) {$x$};
        \node [smallblock, right = 3 em of input,text width=1.6cm] (Neuron) {Neuron};
        % , minimum height=3em, text width=1.6cm
        
        \node[right = 3 em of Neuron] (output) {$y$};
        
        \draw[->] (input) -- (Neuron);
        \draw[->] (Neuron) -- (output);
        \draw[->] (bilinear.north)+(0,0.5) node[xshift = 0em, yshift = +0.5em]{$\theta$} -- (bilinear.north);
        
        % \node at (0,-1.4) {};
    \end{tikzpicture}
    }
    }
    \caption{A neuron.}
    \label{fig:neuron}
\end{figure}

\begin{definition} \rm
A \textit{neuron} is a real-valued function 
\begin{align}
    n \colon \BBR^{l_x} \times \BBR^{l_x+1} \mapsto \BBR
\end{align}
constructed by the composition of a nonlinear function $\sigma \colon \BBR \mapsto \BBR$ and a bi-linear function $\SL \colon \BBR^{l_x} \times \BBR^{l_x+1} \mapsto \BBR,$ that is, $n = \sigma \circ \SL. $
% that is composed of a bi-linear function $\SL \colon \BBR^{l_x} \times \BBR^{l_x+1} \mapsto \BBR$ and a nonlinear function $\sigma \colon \BBR \mapsto \BBR$, that is, $n = \sigma \circ \SL. $
% 
The output of a neuron is thus computed as
\begin{align}
    n(x,\theta)
        &=
            \sigma (\SL(x,\theta) ),
\end{align}
where $\SL(x,\theta) = \matl x^\rmT & 1 \matr \theta.$
Defining $\chi \isdef \matl x^\rmT & 1 \matr^\rmT \in \BBR^{l_x+1}$, it follows that $\SL(x,\theta) = \chi^\rmT \theta.$
The vector $\theta \in \BBR^{l_x+1}$ is called the \textit{neuron gain}. 
The nonlinear function $\sigma$ is also called the \textit{activation function}. 

An example of an activation function is the sigmoid function 
\begin{align}
    \sigma(z)
        =
            \frac{1}
            {1+e^{-z}}.
\end{align}
Alternatively, softmax, linear, harmonic, logarithmic, ReLU, and other functions are also used as the activation function \cite{sharma2017activation}. 
A neuron is shown in Figure \ref{fig:neuron}.
% \begin{align}
%     n:\BBR^{l_x} \times \BBR^{l_x} &\to \BBR,
%     \nn \\
%     n(\theta, x) 
%         &=
%             \frac{1}
%             {1+e^{-\theta^\rmT x}}.
% \end{align}

\end{definition}

A \textit{neural layer} is composed of several neurons. 
The output of the neural layer is a vector whose dimension is equal to the number of neurons in the neural layer. 
Note that each neuron receives the same input.
\begin{definition}
\rm
Let $l_\theta \isdef l_x+1.$
An \textit{$\ell-$dimensional neural layer} is a vector-valued function 
\begin{align}
    N:\BBR^{l_x} \times \BBR^{l_\theta \times \ell} \mapsto  \BBR^{\ell},
\end{align}
constructed by $\ell$ neurons. 
The output of a neural layer is computed as
\begin{align}
    N(x, \Theta) 
        =
            \matl
                n (x, \Theta e_1) \\
                \vdots \\
                n (x,\Theta e_\ell )
            \matr,
\end{align}
% where each neuron $n_i$ is parameterized by the vector $\theta_i \in \BBR^{l_x}.$
where $\Theta \isdef \matl \theta_1 & \ldots & \theta_{\ell} \matr \in \BBR^{l_\theta \times \ell}$ is the \textit{neural layer gain matrix}
and for $i = \{1, \ldots, \ell\},$ $e_i$ is the $i$th column of the $\ell \times \ell$ identity matrix. 
% $I_\ell$
Note that the neural layer 
% is parameterized by the matrix $\Theta \isdef \matl \theta_1 & \ldots \theta_{l_y} \matr \in \BBR^{l_x \times l_y},$ and thus 
can be written as
\begin{align}
    N(x,\Theta) 
        =
            \matl
                \sigma(\SL(x, \Theta e_1) \\
                \vdots \\
                \sigma(\SL(x, \Theta e_{l_y})
            \matr.
\end{align}

\end{definition}

\begin{figure}[h]
    \centering
    \subfloat[Anatomy of a neural layer.]
    {
    \centering
    % \resizebox{\columnwidth}{!}
    {
    \begin{tikzpicture}[auto, node distance=2cm,>=latex',text centered]
        \draw [draw=black, fill = gray, opacity=0.25] (-2.5,-2.75) rectangle (3.25,3.25);
        
        \node at (-3,0) (input) {$x$};
        \node [smallblock, right = 3 em of input,text width=1.6cm] (Neuron) {Neuron $j$};
        
        \node [smallblock, above = 3 em of Neuron,text width=1.6cm] (Neuron_1) {Neuron $1$};
        
        \node [smallblock, below = 3 em of Neuron,text width=1.6cm] (Neuron_ell) {Neuron $\ell$};
        
        \node [smallblock, right = 3 em of Neuron,text width=1.6cm] (Vec) {Vectorize};
        
        % , minimum height=3em, text width=1.6cm

        % \node at {(Plant)+(3,0)}  (output) {$y$};
        \node[right = 3 em of Vec] (output) {$y$};
        
        \draw[->] (input) -- (Neuron);
        \draw[->] (input) -| +(0.75,1) |- (Neuron_1.180);
        \draw[->] (input) -| +(0.75,1) |- (Neuron_ell.180);
        \draw[->] (Neuron) -- (Vec);
        \draw[->] (Neuron_1) -| +(1.5,0) |- (Vec.165);
        \draw[->] (Neuron_ell) -| +(1.5,0) |- (Vec.195);
        \draw[->] (Vec) -- (output);
        % \draw[->] (sum) -- node[xshift = 0em, yshift = .2em]{$e$} (Controller);
        % \draw[->] (Controller)-- node[xshift = 0em, yshift = .2em]{$u$} (Plant);
        % \draw[->] (Plant) -- (output);
        % \draw[->] (Plant.east) -- +(1,0) -- +(+1,-1.5)
        %             -| (sum.south);
        
        \draw[->] (Neuron_1.north)+(0,0.5) node[xshift = -0.75em, yshift = -0.5em]{$\theta_1$} -- (Neuron_1.north);
        \draw[->] (Neuron.north)+(0,0.5) node[xshift = -0.75em, yshift = -0.5em]{$\theta_j$} -- (Neuron.north);
        \draw[->] (Neuron_ell.north)+(0,0.5) node[xshift = -0.75em, yshift = -0.5em]{$\theta_\ell$} -- (Neuron_ell.north);
        
    \end{tikzpicture}
    }
    }
    \\ 
    \subfloat[Simplified representation of a neural layer.]
    {
    \centering
    % \resizebox{\columnwidth}{!}
    {
    \begin{tikzpicture}[auto, node distance=2cm,>=latex',text centered]
        \node at (-3,0) (input) {$x$};
        \node [smallblock, right = 3 em of input,text width=1.6cm] (Neuron) {Neural Layer};
        % , minimum height=3em, text width=1.6cm
        
        \node[right = 3 em of Neuron] (output) {$y$};
        
        \draw[->] (input) -- (Neuron);
        \draw[->] (Neuron) -- (output);
        \draw[->] (bilinear.north)+(0,0.5) node[xshift = 0em, yshift = +0.5em]{$\Theta$} -- (bilinear.north);
        
        % \node at (0,-2.65) {};
    \end{tikzpicture}
    }
    }
    \caption{A neural layer.}
    \label{fig:neural_layer}
\end{figure}
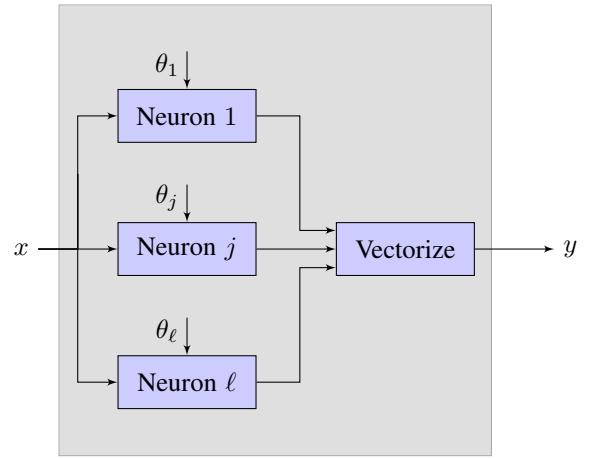
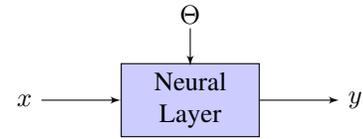

A neural layer is shown in Figure \ref{fig:neural_layer}.

% Note that a neuron layer is parameterized by $l_x l_y$ parameters. 
% Furthermore, a neuron layer with one output is a neuron. 

% \begin{definition}
% \rm
% A \textit{neuron layer} is a function $N:\BBR^{l_x \times l_y} \times \BBR^{l_x} \to \BBR^{l_y}$ such that
% \begin{align}
%     N(\Theta, x) 
%         =
%             \matl{c}
%                 n_1 (\Theta e_1, x) \\
%                 \vdots \\
%                 n_{l_y} (\Theta e_{l_y}, x)
%             \matr.
% \end{align}
% Note that a neuron layer is parameterized by $l_x l_y$ parameters. 
% Furthermore, a neuron layer with one output is a neuron. 

% \end{definition}

A \textit{neural network} is composed of several neural layers. Each neural layer can have an arbitrary number of neurons.

\begin{definition}
\rm 
An \textit{$n$-layer neural network} is a vector-valued function 
\begin{align}
    NN \colon 
\BBR^{l_{x}} \times 
\BBR^{l_{\theta_1} \times \ell_1} \times 
\cdots \times
\BBR^{l_{\theta_n} \times \ell_n} \times 
\BBR^{l_{\ell_n} \times l_y} 
\mapsto \BBR^{l_y},
\end{align}
constructed by the composition of $n$ neural layers, 
where, for each $i \in \{1, 2, \ldots, n\},$ $l_{x_1} = l_x$,  
$l_{x_{i+1}} \isdef \ell_i,$ and 
$l_{\theta_i} \isdef l_{x_i}+1.$
% that is composed of $n$ neural layers,
Note that the $i$th neural layer $N_i$ is $\ell_i$ dimensional  
and the final layer is typically  linear, that is, 
\begin{align}
    NN
        &= 
            S \circ N_n \circ N_{n-1} \circ \ldots \circ N_1,
\end{align}
where $S$ is a linear map. 
The output of the neural network is computed as
\begin{align}
    NN(x, &\Theta_1, \ldots, \Theta_n,\Theta_{n+1})
        \nn \\
        = 
            \Theta_{n+1} &N_n ( N_{n-1} (\ldots ( N_1(x,\Theta_1)),\Theta_{n-1} )\Theta_n ),
\end{align}

% 

% 
% composition of neural layers. 
\end{definition}

% Note that the $i$th layer in the neural network is $N_i.$
The output of a neural network can be computed using a recursive formula as shown below. 
By denoting the input and the output of the $i$th neural layer by $x_i$ and $x_{i+1},$ it follows that, for $i \in \{1, 2, \ldots, n\},$ 
\begin{align}
    x_{i+1} = N_i(x_i, \Theta_i),
\end{align}
where $\Theta_i \in \BBR^{l_{\theta_i} \times \ell_i}.$
Note that $x_1 \isdef x.$
The output of the network is finally given by
\begin{align}
    y = S(x_{n+1}) =  \Theta_{n+1}^\rmT x_{n+1},
\end{align}
where $\Theta_{n+1} \in \BBR^{l_{x_{n+1}} \times l_y}.$
The set of neural layer gain matrices $\{ \Theta_i\}_{i=1}^{n+1}$ parameterizes the neural network and is called the \textit{neural network gain}. 
%% Sections/NN.tex copy pasted above

\begin{figure}
    \centering
    \subfloat[Anatomy of a neural network]
    {
    \resizebox{\linewidth}{!}{
        \begin{tikzpicture}
            \node[clear] (x) {$x$};
            \node[smallblock, right = 2 em of x] (NN1) {$N_1$};
            \node[clear, right = 2 em of NN1] (NN2) {$\cdots$};
            \node[smallblock, right = 2 em of NN2] (NNn) {$N_n$};
            \node[smallblock, right = 2 em of NNn] (sum) {$S$};
            \node[clear, right = 2 em of sum] (y) {$y$};
            % \node[clear, below = 0.1 em of sum] (optional) {(optional)};
            \draw[->] (x) 
            node[xshift = 1.9em, yshift = 0.5em]{$x_1$} 
            -- (NN1);
            \draw[->] (NN1) 
            node[xshift = 2.25em, yshift = 0.5em]{$x_2$} 
            -- (NN2);
            \draw[->] (NN2) node[xshift = 1.6em, yshift = 0.5em]{$x_{n}$} -- (NNn);
            \draw[->] (NNn) 
            node[xshift = 2.5em, yshift = 0.5em]{$x_{n+1}$} 
            -- (sum);
            \draw[->] (sum) -- (y);

            \draw[->] (NN1.north)+(0,0.5) node[xshift = -0.75em, yshift = -0.5em]{$\Theta_1$} -- (NN1.north);
            \draw[->] (NNn.north)+(0,0.5) node[xshift = -0.75em, yshift = -0.5em]{$\Theta_{n}$} -- (NNn.north);
            \draw[->] (sum.north)+(0,0.5) node[xshift = -1.2em, yshift = -0.5em]{$\Theta_{n+1}$} -- (sum.north);
            
        \end{tikzpicture}
    }}
    \\ 
    \subfloat[Simplified representation of a neural network.]
    {
    \centering
    % \resizebox{\columnwidth}{!}
    {
    \begin{tikzpicture}[auto, node distance=2cm,>=latex',text centered]
        \node at (-3,0) (input) {$x$};
        \node [smallblock, right = 3 em of input,text width=1.6cm] (Neuron) {Neural Network};
        % , minimum height=3em, text width=1.6cm
        
        \node[right = 3 em of Neuron] (output) {$y$};
        
        \draw[->] (input) -- (Neuron);
        \draw[->] (Neuron) -- (output);
        \draw[->] (bilinear.north)+(0,0.5) node[xshift = 0em, yshift = +0.5em]{$\{\Theta_i\}_{i=1}^{n+1}$} -- (bilinear.north);
        
        % \node at (0,-2.65) {};
    \end{tikzpicture}
    }
    }
    \caption{A Neural Network.}
    \label{fig:nn_simple}
\end{figure}

\section{Neural Network Training}
\label{sec:NN_training}
% *Need to fix Fig 3 A Neural Network for the sizing and changing $NN$ to $N$*
The objective of training a neural network is to compute a neural network gain such that, for a given input, the neural network's output approximates the correct output. 
The structure of the desired output depends on the application.
For example, in a function approximation application, the output of a trained neural network is expected to closely match the value of the function at the given input. 
In an object classification application, the output of a trained neural network is expected to be a numerical value or a unit canonical vector mapped to the object.

Let $\{(x^1, y^1), \ldots , (x^L, y^L) \}$ denote the training dataset with $L$ elements. 
The neural network is trained by minimizing a cost function of the form
\begin{align}
    J(\{ \Theta_i\}_{i=1}^{n+1})
        =
            \sum_{\ell=1}^L
            \Big( 
                {\| y^\ell - NN(x^\ell, \{ \Theta_i\}_{i=1}^{n+1}) \|_p}
            \Big)^p,
    \label{eq:nn_cost}
\end{align}
where
% $N$ is the number of elements in the training data set, and 
$p$ is a positive integer.   
In most applications, $p=2$, which implies that $J$ is the sum of squares of the prediction errors. 
The training of a neural network is, mathematically, thus the optimization of the cost function \eqref{eq:nn_cost}

The cost function $J$ is generally nonconvex, and, in general, an analytical closed-form solution does not exist  for the minimizer of  \eqref{eq:nn_cost}.
Neural networks are therefore trained using numerical optimization techniques. 
% 

% match the value of the function at the given input. 

In this section, we describe three neural network training methods, namely, the gradient-descent (GD) method, the root-finding method, and the random search method (RSM). 
The gradient-descent method updates the neural network gains in the direction which is opposite to the gradient of the cost function \cite{boyd2004convex}.
In the gradient-descent method, the gradient can be computed either analytically or numerically. 
The root-finding method and the random search method are motivated by the gradient-free numerical optimization techniques such as interior point methods \cite{potra2000interior} and the simulated annealing \cite{press2007numerical}.
% and do not require the computation of the gradient, and thus are gradient-free. 
% gradient-based and novel gradient-free techniques to train a neural network. 
% 

% The gradient can be computed either analytically or numerically. 
% 

\subsection{Gradient-Descent method}
% 
% Defining $\hat y = NN(x, \hat \Theta_1, \ldots, \hat \Theta_n, \hat \Theta_{n+1}),$
% it follows that 
The gradient-based methods compute a minimizer estimate using the recursive relation 
\begin{align}
    \hat \theta_{ij,k+1} 
        =
            \hat \theta_{ij,k}
            -
            \alpha
            \left. \pdt{J}{\theta_{ij}} \right|
            _{\{ \hat \Theta_{i,k}\}_{i=1}^{n+1}},
\end{align}
where, for each 
$j \in \{1, 2, \ldots, n \},$ 
and 
$i  \in \{ 1, 2, \ldots, \ell_j \}, $ $\hat \theta_{ij,k}$ is the minimizer estimate of the $i$th neuron gain in the $j$th layer, that is $\theta_{ij},$ at the $k$th \textit{iteration}, also known as the \textit{epoch}, of the optimization algorithm
and 
$\alpha >0$ is the \textit{learning rate}. 
Note that $\theta_{ij}$ is the $i$th column of $\Theta_j$ 
and $\{ \hat \Theta_{i,k}\}_{i=1}^{n+1}\textbf{}$ is the neural network gain is the minimizer estimate obtained at the $k$th iteration.

Consider the cost function \eqref{eq:nn_cost}, where $p=2.$
The derivative of $J$ with respect to the neuron gain $\theta_{ij}$ is given by
\begin{align}
    \pdt{J}{\theta_{ij}}
        =
            \sum_{m=1}^N
            2 \left(
                y^m - NN(x^m, \{ \Theta_i\}_{i=1}^{n+1}) 
            \right)
            \pdt{NN}{\theta_{ij}} .
    \label{eq:partiat_J}
\end{align}
Note that the neuron gain $\theta_{ij} \in \BBR^{l_{\theta_i}} $ is the $j$th column of the $i$th neural layer gain matrix $\Theta_i,$ that is, $\theta_{ij} =\Theta_i e_j,$ where $j \in \{1,2,\ldots, \ell_i\}.$
It follows from Fact \ref{fact:dy/dtheta} that the derivative \eqref{eq:partiat_J} for $i \leq n$ is then given by
\begin{align}
    \pdt{NN}{\theta_{ij}}            
        &=
            \Theta_{n+1}^\rmT
            \left(
            \prod_{q=i+1}^{n}
            \Lambda(x_{q}, \Theta_{q}) \Theta_{q}^\rmT \SJ_{l_{x_q}}
            \right)
            \sigma_z(z_{ij}) e_j x_i^\rmT
    \label{eq:pdt_NN}
\end{align}
and for $i=n+1,$
\begin{align}
    \pdt{NN}{\theta_{n+1,j}}
        &=
            e_j x_{n+1}^\rmT.
\end{align}
where $e_j$ is the $j$th column of the $\ell_i\times \ell_i$ identity matrix, 
\begin{align}
    z_{ij} \isdef \SL(x_i, \Theta_i e_j),
\end{align}
$\SJ_{l_{x_q}}$ is given by \eqref{eq:SJ_def}, and 
the function $\Lambda$ is given by \eqref{eq:Lambda_def}.
Since the analytical formula to compute the gradient \eqref{eq:partiat_J} requires the computation of \eqref{eq:pdt_NN}, which uses the neural layers gain from the last layer to the $i$th layer, this process is often known as backpropagation
\cite{boden2002guide,goodfellow2016deep}.

\begin{exmp}
    \label{exmp:NN_1_2}
    % \textbf{With bias. }
    Consider a 1-layer neural network with $l_x = 2$ and $l_y = 1.$
    Let $l_{1}  = 2.$
    Then, 
    \begin{align}
        x_1 
            &=
                x \in \BBR^{2},
        \\
        x_2 
            &=
                N_1(x_1, \Theta_1) \in \BBR^{2},
        \\
        y 
            &=
                \Theta_2^\rmT x_2 \in \BBR^{1}.
    \end{align}
    Note that 
    $l_{x_1} = 2,$
    $l_{x_2} = 2$,
    $l_{\theta_1} = 3,$
    $l_{\theta_2} = 2$,
    and thus
    $\Theta_1 \in \BBR^{3 \times 2}$ and 
    $\Theta_2 \in \BBR^{2 \times 1}.$
    Furthermore, 
    \begin{align}
        \pdt{y}{\theta_{21}}
            &=
                x_2^\rmT,
            \\
        % \pdt{y}{\theta_{22}}
        %     &=
        %         e_2 x_2^\rmT,
        %     \\
        \pdt{y}{\theta_{11}}
            &=
            %     \pdt{\Theta_2^\rmT x_2}{\theta_{11}} 
            %     \nn \\
            % &=
            %     \pdt{\Theta_2^\rmT x_2}{x_2} 
            %     \pdt{x_2}{\theta_{11}} 
            %     \nn \\
            % &=
            %     \Theta_2^\rmT
            %     \pdt{N(x_1,\Theta_1)}{\theta_{11}}
            %     \nn \\
            % &=
                \Theta_2^\rmT
                \sigma_z(z_{11}) e_1 \chi_1^\rmT, 
            \\
        \pdt{y}{\theta_{12}}
            &=
            %     \pdt{\Theta_2^\rmT x_2}{\theta_{12}} 
            %     \nn \\
            % &=
            %     \pdt{\Theta_2^\rmT x_2}{x_2} 
            %     \pdt{x_2}{\theta_{12}} 
            %     \nn \\
            % &=
            %     \Theta_2^\rmT
            %     \pdt{N(x_1,\Theta_1)}{\theta_{12}} 
            %     \nn \\
            % &=
                \Theta_2^\rmT
                \sigma_z(z_{12}) e_2 \chi_1^\rmT,
    \end{align}
    where $\chi_1 = \matl x_1^\rmT & 1 \matr.$
\end{exmp}

\subsection{System of Nonlinear Equations}
The objective of minimizing the cost function \eqref{eq:nn_cost} is to find a neural network gain set $\{ \Theta_i\}_{i=1}^{n+1}$ such that
\begin{align}
    NN(x^m, \{ \Theta_i\}_{i=1}^{n+1}) 
        -
            y^m
        =
            0, 
    \label{eq:fsolve_eqn}
\end{align}
where, for $m\in \{1, \ldots, L \}$ $(x^m, y^m)$ are the elements of the training dataset. 
Note that \eqref{eq:fsolve_eqn} is a system of $L$ nonlinear equations, where $\{ \Theta_i\}_{i=1}^{n+1}$ is the unknown parameter. 
The problem of training the neural network can thus be interpreted as the problem of solving the system of nonlinear equations \eqref{eq:fsolve_eqn}. 
In this work, we use the \href{https://www.mathworks.com/help/optim/ug/fsolve.html}{\texttt{fsolve}} (FS) routine in MATLAB to solve \eqref{eq:fsolve_eqn} \cite{fsolve}.

Note that it is assumed that there exists a neural network gain set $\{ \Theta_i\}_{i=1}^{n+1}$ such that \eqref{eq:fsolve_eqn} is satisfied. 
This assumption is not restrictive in practice since the structure of the neural network can be expanded to increase the number of free parameters so as to render \eqref{eq:fsolve_eqn} well-posed.

% MATLAB's \gls{fs} uses the \gls{trd} algorithm
% by default [source: matlab fsolve].
% The \gls{trd} generally minimizes a function 
% $F:\mathbb{R}^n \mapsto \mathbb{R}$.
% The vector inputs $x$ that produce a lower value
% of $F$ is obtained by the equation 
% [source: matlab equation solving algorithm]
% \begin{align}
%     J(x_k) d_k 
%         &= -F(x_k) \\
%     x_{k+1}
%         &= x_k + d_k \\
%     \ni J(x_k)
%         &= 
%             \begin{bmatrix}
%                 \nabla F_1 (x_k)^T \\
%                 \nabla F_2 (x_k)^T \\
%                 \vdots \\
%                 \nabla F_n (x_k)^T
%             \end{bmatrix}
% \end{align}
% When the Jacobian is non-invertible,
% Trust-Region techniques are implemented
% [source: unconstrained nonlinear optimization].

% Consider a series of normalized random gains
% in an arbitrary network $N$

% \begin{equation}
%     J_{[i]} = 
%         \sum \left( y - \hat{y}_{[i]} \right)^2 | 
%             \hat{y}_{[i]} = 
%                 N(x, \Theta_{[i]})
% \end{equation}

% Gains are updated following a $\Theta_{[k]} | \min{J}$

% \begin{equation}
%     \Theta = \Theta_{[k]} + \mathcal{N}(0,1)
% \end{equation}

\subsection{Random Search Method}

The gradient computation using \eqref{eq:partiat_J} in neural network training is typically the most computationally expensive step. 
% 
% The gradient computed using \eqre
% 
To reduce the computational cost and the programming effort required to compute the gradient, we investigate the effectiveness of an admittedly primitive random search optimization method to train the neural network. 
The random search method, described below, is motivated by the simulated annealing used in numerical optimization. 

The neural network is initialized with a set of neural network gains $\{ \Theta_{i,0}\}_{i=1}^{n+1}.$
Note that $\{ \Theta_{i,0}\}_{i=1}^{n+1}$ is the minimizer estimate at the 0th epoch. 
At $k$th epoch, an ensemble of $M$ neural network gains is generated by sampling a hypersphere of radius $\alpha>0$ centered at the neural gain estimates at the $k$th epoch, that is,  
for each $m \in  \{1, \ldots , M\},$
\begin{align}
    \theta_{ij}^m 
        = 
            \theta_{ij,k} + \alpha \nu, 
    \label{eq:RSM_theta_update}
\end{align}
where $\nu \in \SN(0, I_{\ell_i})$ for each
$i \in \{1, \ldots, n\}$
and 
$j \in \{1, \ldots, \ell_i\}.$
Note that $\theta_{ij,k}$ is the $j$th neuron gain estimate in the $i$th neural layer obtained at the $k$th epoch and $\theta_{ij}^m $ is the corresponding gain generated to minimize the cost \eqref{eq:nn_cost}.
The cost $J_k^m$ is computed for each ensemble member. 
The neural network gain estimate $\{ \Theta_{i,k+1}\}_{i=1}^{n+1}$ is finally given by 
\begin{align}
    \{ \Theta_{i,k+1}\}_{i=1}^{n+1}
        =
            \underset{m \in \{1, \ldots, M\} }{\mathrm{argmin}} 
            J_k^m(\{ \Theta_{i,k}^m\}_{i=1}^{n+1}).
\end{align}

\section{Numerical Examples}
\label{sec:examples}
% \subsection{XOR function}
This section applies the three training techniques described in the previous section and compares their convergence rate and prediction accuracy using numerical examples. 

\begin{exmp}
    \label{exmp:xor}
    \textbf{[XOR approximation.]}
    In this example, the objective is to approximate the output of the XOR function using a neural network. 
    XOR function is a discreet function with two inputs and one output, and its values are given in Table \ref{tab:xor_table}.
    We use a 1-layer neural network with a 2-dimensional neural layer, where the activation function is chosen to be the sigmoid function.
    Note that this neural network architecture is shown in Example \ref{exmp:NN_1_2}.
    \begin{table}[h]
        \centering
        \begin{tabular}{|c|c|}
        \hline
            $x$ & $y$
        \\ \hline
            $\matl 0 & 0 \matr$ & $0$
        \\ \hline
            $\matl 0 & 1 \matr$ & $1$
        \\ \hline
            $\matl 1 & 0 \matr$ & $1$
        \\ \hline
            $\matl 1 & 1 \matr$ & $0$
        \\ \hline
        \end{tabular}
        \caption{XOR function}
        \label{tab:xor_table}
    \end{table}

    In the gradient-descent method, we set the learning rate $\alpha = 5.$
    In MATLAB's fsolve routine, the function and step tolerance are set to $1\rme -30.$
    In the random search method, we generate a $50$-member ensemble at each iteration and set the hypersphere radius $\alpha = 1$.     
    Figure \ref{fig:0101_1L_2N_Sigmoid} shows the cost \eqref{eq:nn_cost} with the neural network gains optimized by each training method.
    Table \ref{tab:0102_predictionAccuracy} shows the predicted output of each neural network with the gains obtained at the 50th iteration.
    Finally, Figure \ref{fig:0107_1L_2N_Sigmoid_XOR_Theta} shows the neural network gains after each iteration of training with all three methods.
    Note that the fsolve routine, which solves the nonlinear system of equations \eqref{eq:fsolve_eqn}, outperforms the other two training methods in terms of convergence rate and prediction accuracy. 
    % Finally, in RCPE, we set .....
    \hfill{\huge$\diamond$}
    \begin{figure}[h]
        \centering
        \includegraphics[width=\columnwidth]{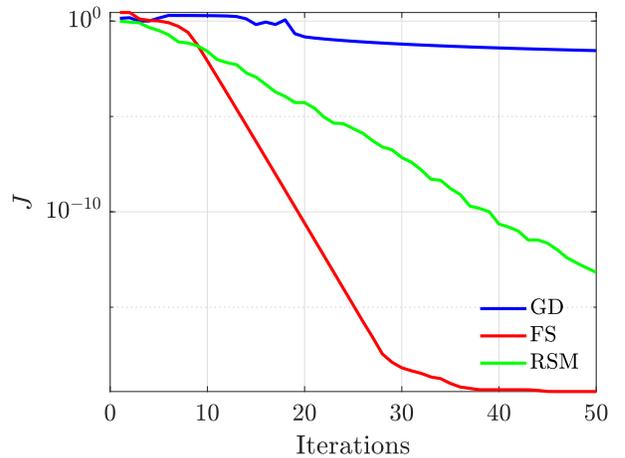}
        \caption{
        \textbf{XOR approximation.} 
        Cost \eqref{eq:nn_cost} computed with the neural network gains optimized by each training method.
        % Note that the convergence rate of the fsolve-based method is 
        }
        \label{fig:0101_1L_2N_Sigmoid}
    \end{figure}

    \begin{table}[h]
        \centering
        \begin{tabular}{|c|c|c|c|}
            \hline
            $x$ & $\hat{y}_{\rm GD}$ & $\hat{y}_{\rm FS}$ & $\hat{y}_{\rm RSM}$ 
            \\
            \hline
            $\matl 0 & 0 \matr$ & $9.7155\rme -02$ & $9.3003\rme -11$ & $1.3148\rme -07$ 
            \\ \hline
            $\matl 0 & 1 \matr$ & $9.1980\rme -01$ & $1.0000\rme +00$ & $1.0000\rme +00$ 
            \\ \hline
            $\matl 1 & 0 \matr$ & $9.2056\rme -01$ & $1.0000\rme +00$ & $1.0000\rme +00$ 
            \\ \hline
            $\matl 1 & 1 \matr$ & $7.7451\rme -02$ & $6.5421\rme -11$ & $1.5650\rme -07$
            \\
            \hline
        \end{tabular}
        \caption{\textbf{XOR approximation.} 
        {Neural network predictions at the 50th iteration.}}
        \label{tab:0102_predictionAccuracy}
    \end{table}
    \begin{figure}[h]
        \centering
        \includegraphics[width=\columnwidth]{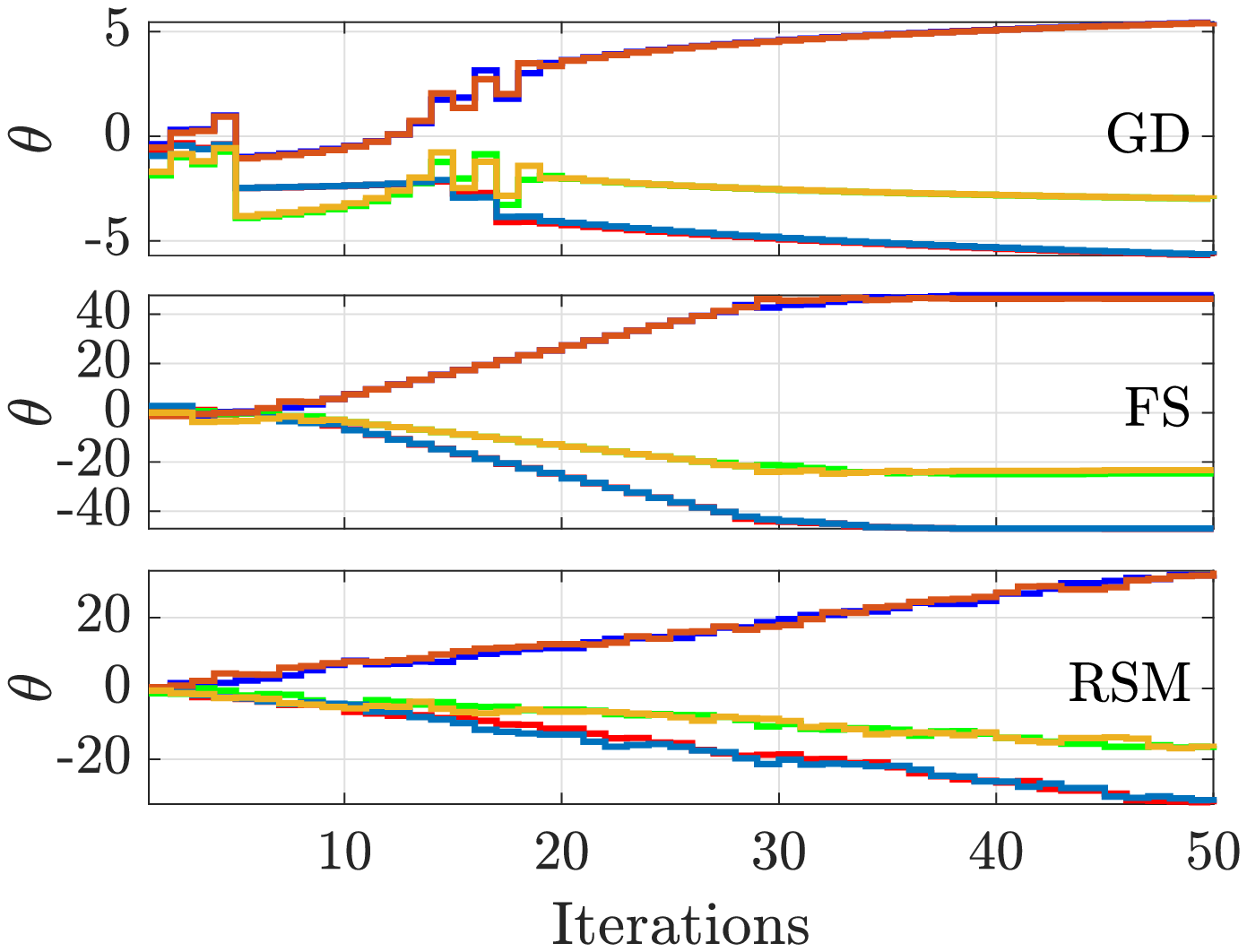}
        \caption{\textbf{XOR approximation.} 
        Neural network gains optimized by each training method.
        }
        \label{fig:0107_1L_2N_Sigmoid_XOR_Theta}
    \end{figure}

% 
    % Figure 
    % neural network with 1 layer, 2 neurons, and sigmoid function. 
    % show convergence results for each method.
    % show prediction accuracy. 
\end{exmp}

% Note that the methods in figure \ref{fig:0101_1L_2N_Sigmoid}
% could be tuned.
% In particular, a constant learning rate for the \gls{gd}
% was set to 5; 
% the function and step tolerance, and the maximum function evaluations,
% were set to $1\times 10^{30}$ and $1\times 10^{20}$ respectively
% for \gls{fs};
% 50 samples of searches were used in each iteration for the \gls{rsm}.
% Additionally, \gls{fs} stops prematurely upon reaching the specific tolerances or evaluations.

% The structure for thin neural network can be mathematically written as:
% \begin{align}
%     y       &= \begin{bmatrix}
%                     1\\
%                     1
%                 \end{bmatrix}^T
%                 x_2 \ 
%     &|\ x_2  &= \begin{bmatrix}
%                     x_{2,1} \\
%                     x_{2,2} \\
%                 \end{bmatrix}\nn \\
%     x_{2,i} &= \sigma \left(\chi_1^T \Theta_1 e_i\right) \label{eq:XOR_x2}\ 
%     &| \ \Theta_1 &\in \mathbb{R}^{3\times 2}, \\
%     && \ \chi_1 &= \begin{bmatrix}
%         x_1 \\
%         1
%     \end{bmatrix}, \nn \\
%     && \ x_1 &= x \nn
% \end{align}
% where $e_i$ is the standard basis vector with 1 in the $i$th position,
% which selects a particular column of $\Theta$.
% Mathematically, non-linear functions like the $\sigma$ accept
% scalars by default.
% However, in computational practice, \ref{eq:XOR_x2}
% can be written as $x_2 = \sigma \left(\chi_1^T \Theta_1\right)$

%  
\begin{exmp}
    \textbf{[Trigonometric function approximation.]}
    % neural network with 3 layer, 2 neurons, and sigmoid function. 
    % show convergence results for each method.
    % show prediction accuracy. 
    In this example, we train a neural network to approximate the sine function. 
    Specifically, we consider a 2-layer neural network, with one neuron in each layer, that is,  
    % Both layers were chosen to be one-dimensional with a sigmoid activation function in the first layer.
    % 
    \begin{align}
        x_1 
            &=
                x \in \BBR,
        \\
        x_2 
            &=
                N_1(x_1, \Theta_1) \in \BBR,
        \\
        y 
            &=
                N_2(x_2, \Theta_2) \in \BBR.
    \end{align}
    Note that 
    $l_{x_1}  = l_{x_2} =  l_y = 1$, and thus
    $l_{\theta_1} = 2,$
    $l_{\theta_2} = 2,$
    $\Theta_1 \in \BBR^{2 },$ and 
    $\Theta_2 \in \BBR^{2 }.$
    
    The training data is generated by selecting 100 linearly spaced values between $-\pi/2$ and $\pi/2.$
    % 
% The objective for the second example is to model the sine function
% for a linearly spaced input ranging $[-\frac{\pi}{2}, \frac{\pi}{2}]$,
% using a neural network
% comprised of two layers only, with only one neuron in each.
% An activation function, sigmoid, was used only for the first layer,
% and none was chosen in the second.
% 
In gradient-descent method, we set learning rate $\alpha=0.01.$
In MATLAB's fsolve routine, the function and step tolerance are set to $1\rme -30.$ 
In the random search method, we generate a 500-member ensemble at each iteration and set the hypersphere radius $\alpha = 1$. 
Figure \ref{fig:0103_2L_1N_Sine} shows the cost \eqref{eq:nn_cost} with the neural network gains optimized by each training method.

\begin{figure}[h]
    \centering
    \includegraphics[width=\linewidth]{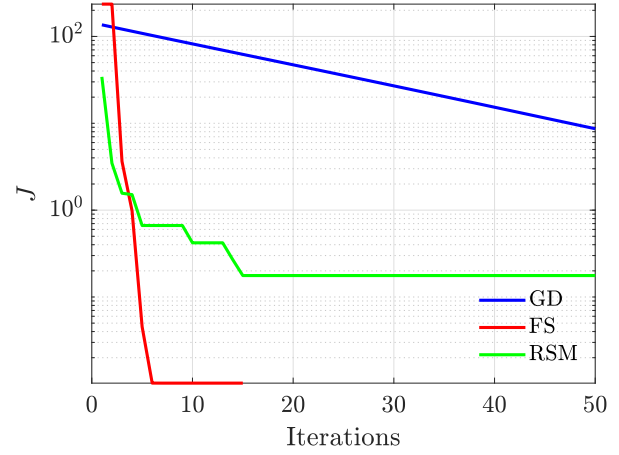}
    \caption{
    \textbf{Sine approximation.} 
        Cost \eqref{eq:nn_cost} computed with the neural network gains optimized by each training method.}
    \label{fig:0103_2L_1N_Sine}
\end{figure}

% with a decay factor of $\delta = 0.98$.
% In each epoch $k$, we generate 500 searches with a decreasing radius of
% $\alpha_k = \delta \cdot \alpha_{k-1}$.
% We used the same function tolerance, step tolerance, and maximum function evaluations as in the XOR example.

% Figure \ref{fig:0103_2L_1N_Sine} shows the cost \eqref{eq:nn_cost} computed with the neural network gains optimized by each training method after each iteration. 
% 
To investigate the approximation accuracy of the three trained neural network, we compute the error $|y-\hat y|, $ where $y= \sin (x)$ and $\hat y = NN(x, (\Theta_1, \Theta_2)),$ at randomly generated values of $x$ between $-\pi/2$ and $\pi/2.$
% neural network results. 
Figure \ref{fig:0105_2L_1N_Sine_Error} shows the error with the neural network gains obtained at the 50th iteration. 
Figure \ref{fig:0108_2L_1N_Sine_Theta} shows the neural network gains after each iteration of training with all three methods.
Similar to the previous example, the fsolve routine, which solves the nonlinear system of equations \eqref{eq:fsolve_eqn}, outperforms the other two training methods in terms of convergence rate and prediction accuracy. 
\hfill{\huge$\diamond$}

% Figure \ref{fig:0104_2L_1N_Sine} illustrates the decrease in cost
% using the neural network gains for different training methods.
% Figure \ref{fig:0105_2L_1N_Sine_Error} computes the error between
% the actual sine function and the neural network generatesd sine function
% for uniformly generated random input values in $[-\frac{\pi}{2}, \frac{\pi}{2}]$.

% \begin{figure}[htbp]
%     \centering
%     \includegraphics[width=\linewidth]{Figures/RSCH_0104_2Layer_1Neurons_1Sigmoid_2Linear_Sine_Y.eps}
%     \caption{neural network Based Sine Function for Limited Domain}
%     \label{fig:0104_2L_1N_Sine}
% \end{figure}

\begin{figure}[h]
    \centering
    \includegraphics[width=\linewidth]{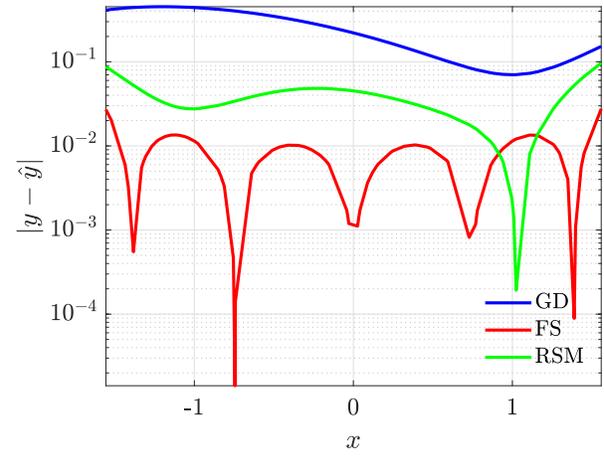}
    \caption{\textbf{Sine approximation.} Neural network prediction error at the 50th iteration.}
    \label{fig:0105_2L_1N_Sine_Error}
\end{figure}

\begin{figure}[h]
    \centering
    \includegraphics[width=\linewidth]{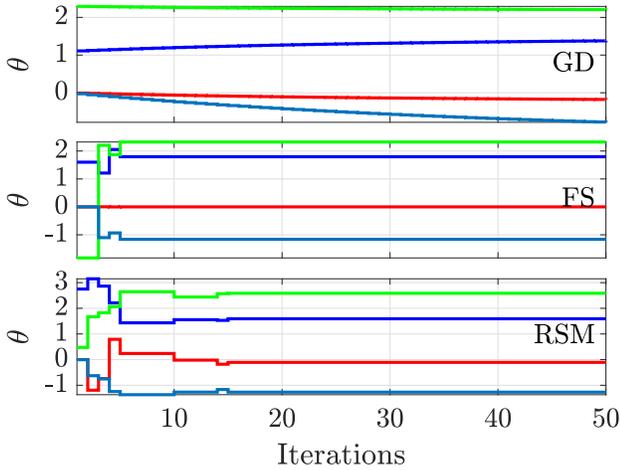}
    \caption{\textbf{Sine approximation.} Neural network gains optimized by each training method.}
    \label{fig:0108_2L_1N_Sine_Theta}
\end{figure}

\end{exmp}

\begin{exmp}
    \textbf{[Handwritten digit identification.]}
    In this example, we train a neural network to identify handwritten numbers.
    To keep computational requirements low, we consider the problem of identifying the integers $0,$ $1,$ and $2.$
    We consider a 2-layer neural network
    with 30 neurons in the first layer and 3 neurons in the second layer.
    The activation functions are chosen to be ReLU in the first layer and sigmoid in the second layer. 
    Therefore,
    \begin{align}
        x_1 
            &=
                x \in \BBR^{784},
        \\
        x_2 
            &=
                N_1(x_1, \Theta_1) \in \BBR^{30},
        \\
        y 
            &=
                N_2(x_2, \Theta_2) \in \BBR^{3}.
    \end{align}
    Note that 
    $l_{x_1}  = 784,$
    $l_{x_2} =  30,$
    $l_y = 3$, and thus
    $l_{\theta_1} = 785,$
    $l_{\theta_2} = 31,$
    $\Theta_1 \in \BBR^{785 \times 30},$ and 
    $\Theta_2 \in \BBR^{31\times 3}.$
    
    % The training data is obtained by extracting the first 30 images and labels
    % of the MNIST dataset that correspond to the digits 0, 1, and 2,
    % using a \href{https://github.com/amaas/stanford_dl_ex/tree/master/common}{pre-existing import script}\footnote{\url{https://github.com/amaas/stanford_dl_ex/tree/master/common}}.
    % Note that the ReLU function is given by
    % \begin{equation}\label{eq:relu}
    %     \mathrm{ReLU}(x) = \begin{cases}
    %         x & \text{if }x \ge 0 \\
    %         0 & \text{if }x < 0
    %     \end{cases}
    % \end{equation}

    The training data set consists of 30 labeled images corresponding to the digits $0,$ $1,$ and $2$ in the MNIST data set.
    We use the gradient-descent method implemented in TensorFlow with a learning rate of  $\alpha = 0.7$.
    % 
    % In the gradient-descent method, TensorFlow is used with a learning rate of $\alpha = 0.7$.
    In MATLAB's fsolve, the default optimization options are used.
    In the random search method, we generate a 5000-member ensemble at each iteration and set the hypersphere radius $\alpha = 1$.
    Figure \ref{fig:0106_2L_MNIST_Cost} shows the cost \eqref{eq:nn_cost} with the neural network gains optimized by each training method.
    Note that the cost $J$ drops to exactly $0$ at the 13th iteration in the random search method, which is rare in typical parameter fitting problems, but is plausible due to a small number of training samples and a large number of parameters being fitted.  
    
    To determine the accuracy of the trained neural network,
    we use the three trained neural networks to identify 100 samples of digits 0, 1, and 2.
    Note that we use the three trained neural networks to identify digits from a validation dataset that is different from the training dataset. 
    Table \ref{tab:MNIST} shows the neural network predictions of sample handwritten digits.
    %     
    % example predictions for three digits using the trained neural network.
    % 
    Figure \ref{fig:0109_2L_MNIST_Bar} shows the number of correct identifications for each of the digits predicted by the three networks.
    Note that, unlike the previous examples, the neural network trained by the random search method outperforms the other two networks. 
    \hfill{\huge$\diamond$}
    
    % The training data was selected as a subset of the MNIST dataset.
    % The MNIST dataset is a collection of 70000 labeled handwritten integers
    % ranging from 0 to 9, imported using a \href{}{preexisting script}\footnote{\url{https://github.com/amaas/stanford_dl_ex/tree/master/common}}.
    % For the purpose of this training, only the first
    % 30 training numbers ranging from 0 to 2 were used.
    % The first layer had 30 neurons with ReLU as the activation function.
    
    % The second layer had 3 neurons with the sigmoid activation function.
    % % The cost was given by
    % % \begin{equation}
    % % J = \frac{1}{n} \sum_{i=1}^{n} (y_i-\hat{y}_i)^T (y_i-\hat{y}_i)
    % % \end{equation}
    % TensorFlow, a modern machine learning platform, was used to conduct the gradient-based training
    % with $\alpha=5$.
    % In RSM, for a starting $\alpha = 1$, we generated $5000$-member ensemble in each iteration.
    % In FSolve, the default settings were used.
    % The iterations had to be decreased to 20 since the calculations were extremely time-consuming for FSolve.

    % Figure \ref{fig:0106_2L_MNIST_Cost} shows the decrease in cost for the three methods for MNIST.
    % Figure \ref{fig:0109_2L_MNIST_Bar} depicts the number of correct predictions
    % for 30 training and novel testing images.

    \begin{figure}[h]
        \centering
        \includegraphics[width=\linewidth]{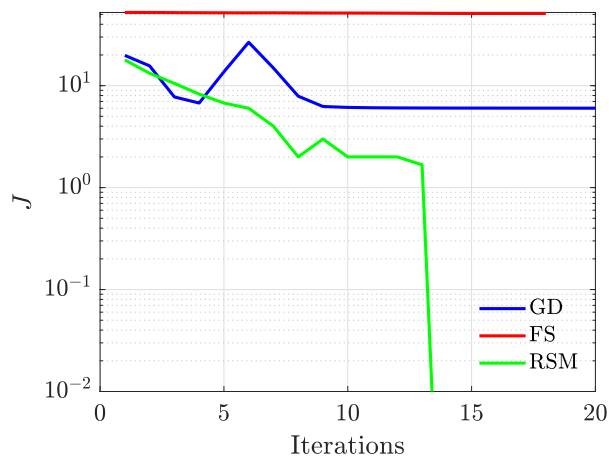}
        \caption{\textbf{Digit identification.} Cost \eqref{eq:nn_cost} computed with the neural network gains optimized by each training method.}
        \label{fig:0106_2L_MNIST_Cost}
    \end{figure}

    \begin{table}[h]
        \centering
        \begin{tabular}{|c|c|c|c|c|}
            \hline
            $x$    & $y$ & $\hat{y}_\mathrm{GD}$    & $\hat{y}_\mathrm{FS}$     & $\hat{y}_\mathrm{RSM}$
            \\
            \hline
             \includegraphics[width=0.1\textwidth,valign=m]{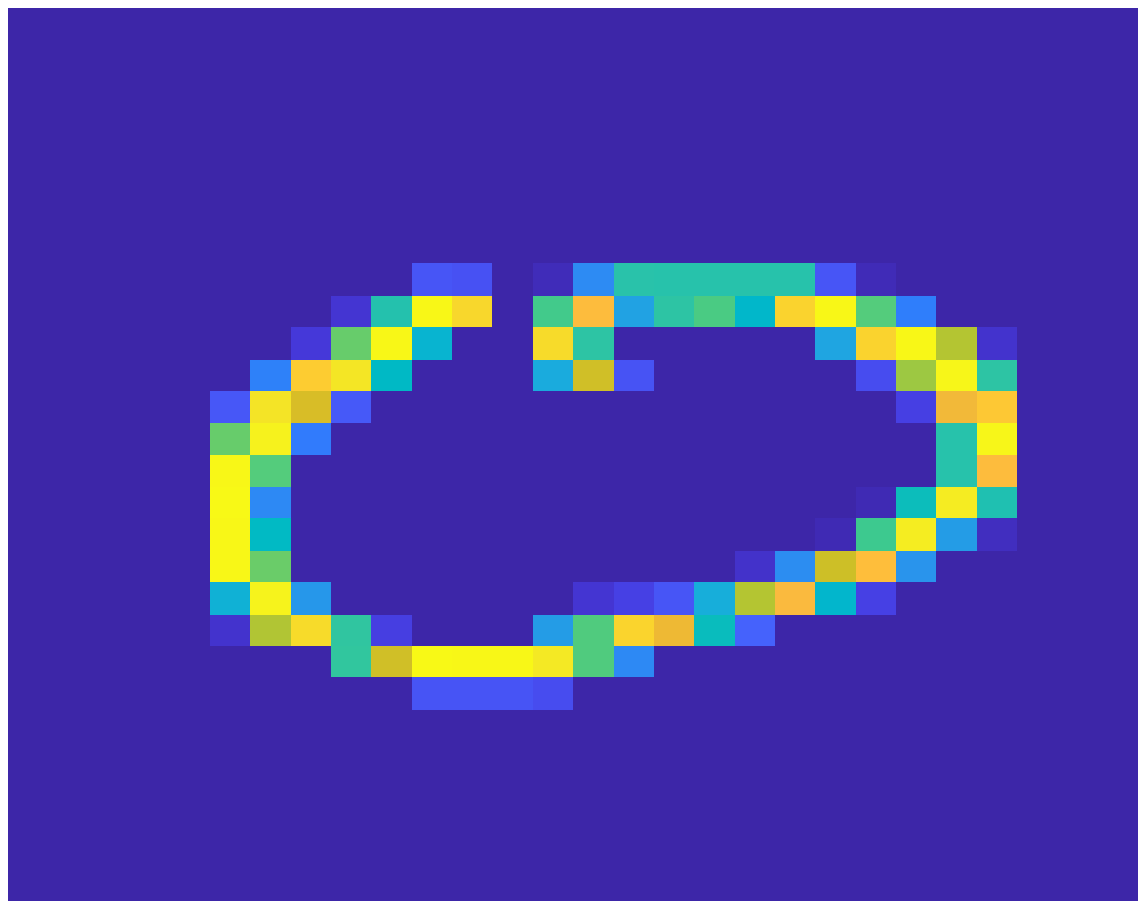}
             & $\begin{bmatrix}
                 1 \\
                 0 \\
                 0
             \end{bmatrix}$     &   $\begin{bmatrix}
                        0.152 \\
                        0.139 \\
                        0.149
                        \end{bmatrix}$        & $\begin{bmatrix}
                                                1.000 \\
                                                0.000 \\
                                                1
                                                \end{bmatrix}$          &   $\begin{bmatrix}
                                                                            1 \\
                                                                            0.000 \\
                                                                            0.000 \\
                                                                            \end{bmatrix}$
            \\ \hline
            \includegraphics[width=0.1\textwidth,valign=m]{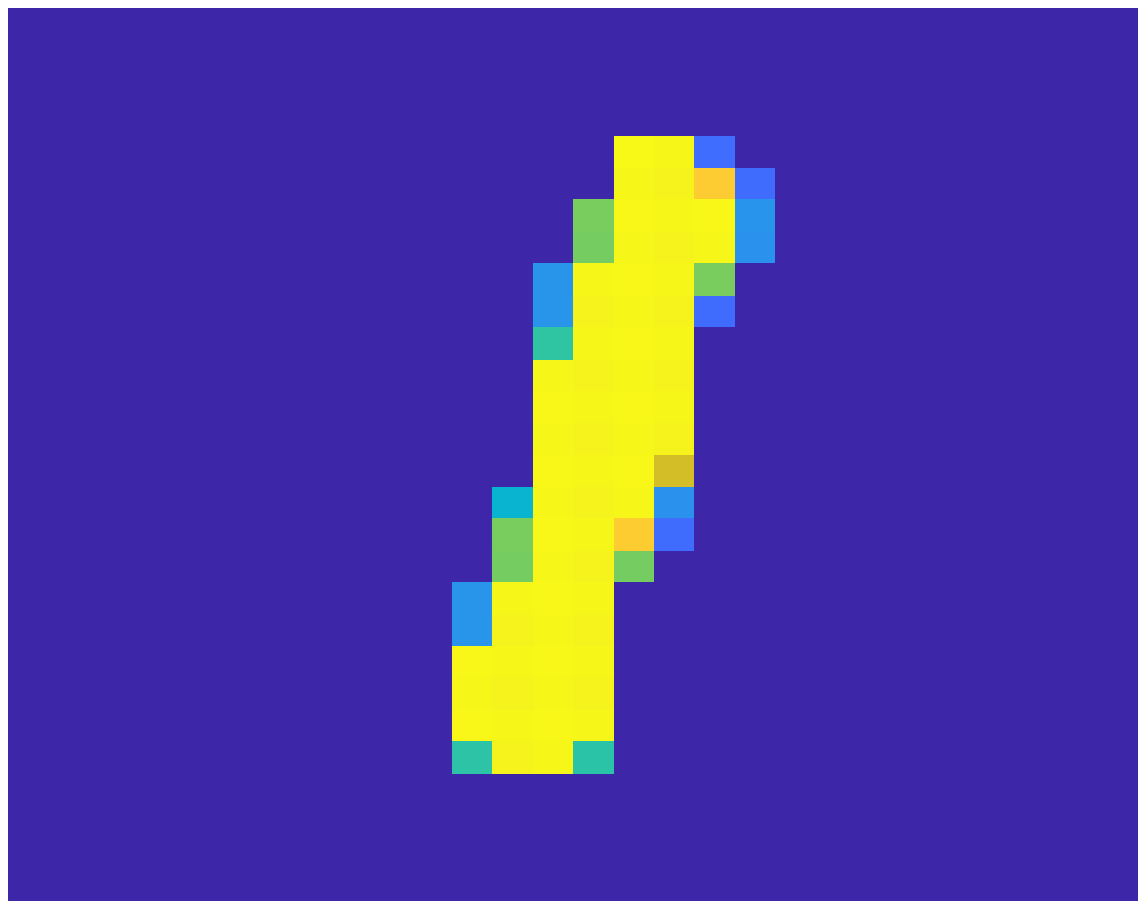}
            & $\begin{bmatrix}
                 0 \\
                 1 \\
                 0
             \end{bmatrix}$        & $ \begin{bmatrix}
                        1.000 \\
                        1.000 \\
                        1.000 \\
                        \end{bmatrix}$        & $   \begin{bmatrix}
                                                    1\\
                                                    1.000 \\
                                                    0.986
                                                    \end{bmatrix}$      & $ \begin{bmatrix}
                                                                            0 \\
                                                                            1 \\
                                                                            0.000
                                                                            \end{bmatrix}$
            \\ \hline 
            \includegraphics[width=0.1\textwidth,valign=m]{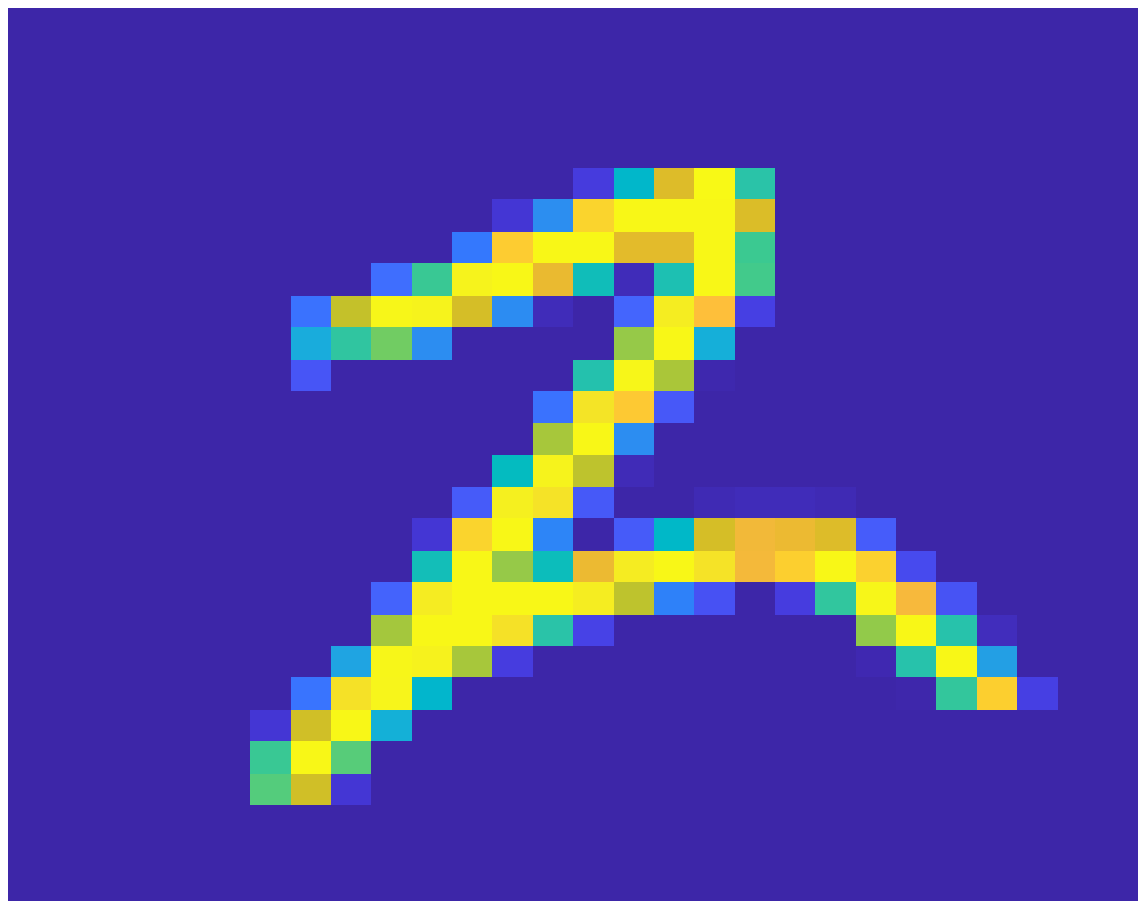}
            & $\begin{bmatrix}
                 0 \\
                 0 \\
                 1
             \end{bmatrix}$       & $ \begin{bmatrix}
                        1.000 \\
                        1.000 \\
                        1.000
                        \end{bmatrix}$          & $ \begin{bmatrix}
                                                    1 \\
                                                    0.015 \\
                                                    1.000
                                                    \end{bmatrix}$      & $ \begin{bmatrix}
                                                                            0.000 \\
                                                                            0.000 \\
                                                                            1
                                                                            \end{bmatrix}$
            \\
            \hline
        \end{tabular}
        \caption{\textbf{Digit identification.} Neural network predictions of sample handwritten digits. 
        Note that the digits are sampled from a validation dataset that is different from the training dataset. 
        }
        \label{tab:MNIST}
    \end{table}

    \begin{figure}[h]
        \centering
        \includegraphics[width=\linewidth]{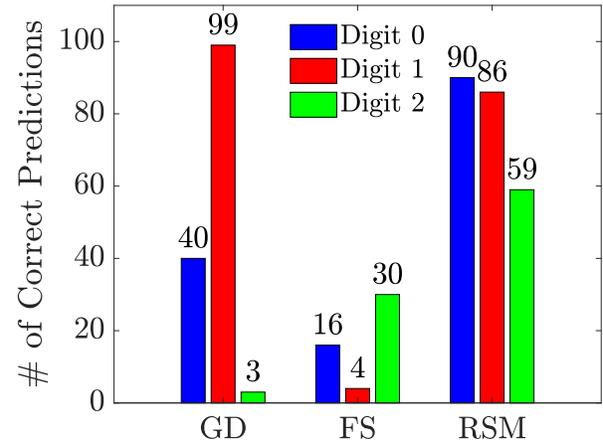}
        \caption{\textbf{Digit identification.} 
        The number of correct predictions made by the three trained neural networks out of 100 validation samples of each digit. 
        % Neural network accuracy on training and testing MNIST data}
        }
        \label{fig:0109_2L_MNIST_Bar}
    \end{figure}
    
\end{exmp}

\section{Conclusion}
\label{sec:conclusion}

This paper presented a compact, matrix-based representation of neural networks in a self-contained tutorial fashion.
The neural networks are represented as a composition of several nonlinear mathematical functions. 
Three numerical methods, one based on gradient-descent and two gradient-free, are reviewed for training a neural network.
The reviewed training methods are applied to three typical machine learning problems.
Surprisingly, the neural networks trained with the gradient-free methods outperformed the neural network trained with the widely used gradient-based optimizer. 
 
\printbibliography

\section*{Useful Facts}
%% Sections/UsefulFacts.tex copy pasted below
% \subsection{Useful Facts}
The following facts review the well-known results from multivariable calculus. 
\begin{fact}
    \textbf{Derivatives of scalar functions with respect to vectors.}
    Let $a \in \BBR^n,$ $A \in \BBR^{n \times n},$ and $f \colon \BBR^n \to \BBR.$
    Let $x \in \BBR^n$.
    Then, 
    \begin{align}
        \pdt{x^\rmT a}{x}
            &=   
                \pdt{a^\rmT x}{x}
            =   
                a^\rmT \in \BBR^{1 \times n}
            , \\
        \pdt{x^\rmT A x}{x}
            &=   
                2x^\rmT A \in \BBR^{1 \times n},
        \\
        \pdt{f}{x}
            &=   
                \matl 
                    \pdt{f}{x_1} &
                    \cdots &
                    \pdt{f}{x_n} 
                \matr
                 \in \BBR^{1 \times n}.
    \end{align}
\end{fact}
\begin{fact}
    \textbf{Derivatives of vector functions with respect to vectors.}
    Let $A \in \BBR^{m \times n}$ and $f \colon \BBR^n \to \BBR^m.$
    Let $x \in \BBR^n$
    Then, 
    \begin{align}
        \pdt{A x}{x}
            &=   
                A \in \BBR^{m \times n},
        \\
        \pdt{f}{x}
            &=   
                \matl 
                    \pdt{f_1}{x_1} &
                    \cdots &
                    \pdt{f_1}{x_n} 
                    \\ 
                    \vdots & \ddots & \vdots \\
                    \pdt{f_m}{x_1} &
                    \cdots &
                    \pdt{f_m}{x_n} 
                \matr
                 \in \BBR^{m \times n}.
    \end{align}
\end{fact}

\begin{fact}
    \textbf{Derivatives of scalar functions with respect to matrices.}
    Let $a \in \BBR^n,$
    $b \in \BBR^m,$
    $A \in \BBR^{n \times n},$ and
    $f \colon \BBR^{n\times m} \to \BBR.$
    Let $X \in \BBR^{n \times m}$.
    Then, 
    \begin{align}
        \pdt{a^\rmT X b}{X}
            &=   
                b a^\rmT \in \BBR^{m \times n}
            , \\
        \pdt{f}{X}
            &=   
                \matl 
                    \pdt{f}{X_{11}} &
                    \cdots &
                    \pdt{f}{X_{n1}} \\ 
                    \vdots & \ddots & \vdots \\
                    \pdt{f}{X_{1m}} &
                    \cdots &
                    \pdt{f}{X_{mn}} 
                \matr
                 \in \BBR^{m \times n}.
    \end{align}
\end{fact}
% \begin{fact}
%     Useful matrix facts.
%     Let $a,x \in \BBR^n, $ $b \in \BBR^m,$ 
%     $A \in \BBR^{n \times n}$
%     and
%     $B \in \BBR^{m \times n}.$
%     Let $X \in \BBR^{m \times m}.$
%     Then, 
%     \begin{align}
%         \pdt{x^\rmT a}{x}
%             &=   
%                 \pdt{a^\rmT x}{x}
%             =   
%                 a^\rmT \in \BBR^{1 \times n}
%             , \\
%         \pdt{x^\rmT A x}{x}
%             &=   
%                 2x^\rmT A \in \BBR^{1 \times n}
%             , \\
%         \pdt{B x}{x}
%             &=   
%                 B \in \BBR^{m \times n}
%             , \\
%     \end{align}
% \end{fact}

\begin{fact}
    The derivative of the sigmoid function $\sigma(z) = \dfrac{1}{1+e^{-z}}$ is given by
    \begin{align}
        \sigma_z(z)
            \isdef 
        \pdt{\sigma}{z} 
            = 
                \sigma(z) [1-\sigma(z)].
    \end{align}
\end{fact}

\begin{fact}
    \textbf{Chain rule. }
    Let $f \colon \BBR^m \to \BBR^p$ and 
    % $g \colon \BBR^n \to \BBR^m$.
    Define $h \colon \BBR^n \to \BBR^p $ as the composition of $f$ and $g$ , that is, $ f \circ g,$ or, $h(x) = f(g(x)),$ where 
    $x \in \BBR^{n}.$
    Then, 
    \begin{align}
        \pdt{h}{x}
            = 
                \pdt{h}{y} \pdt{y}{x},
    \end{align}
    where $y \isdef g(x) \in \BBR^m.$
    Note that 
    $\pdt{y}{x} \in \BBR^{m \times n}$ and
    $\pdt{h}{y} \in \BBR^{p \times m}$, and thus
    $\pdt{h}{x} \in \BBR^{p \times n}.$
\end{fact}

The following results apply well-known multivariable calculus results to multivariable functions typically used in neural network training. 
\begin{fact}
    Let $x\in \BBR^{l_x}$ and define $\chi \isdef \matl x^\rmT & 1 \matr^\rmT.$
    Then,
    \begin{align}
        \pdt{\chi}{x}
            =
                \SJ_{l_x},
    \end{align}
    where 
    \begin{align}
        \SJ_{l_x} \isdef \matl 
            I_{l_x} \\ 0_{1 \times {l_x} }
        \matr
        \in \BBR^{l_x+1 \times l_x}.
        \label{eq:SJ_def}
    \end{align}
\end{fact}

\begin{fact} 
    Let $z \isdef \SL(x, \theta) = \chi^\rmT \theta,$
    where $\theta \in \BBR^{l_\theta}.$
    Note that $l_\theta = l_x+1.$
    Let $\sigma \colon \BBR \to \BBR.$
    Then, 
    \begin{align}
        \pdt{\sigma(\SL(x, \theta)}{{\theta}}
            &=
                \pdt{\sigma}{{z}}
                \pdt{z}{{\theta}}
            =
                \sigma_z(z)
                \chi^\rmT
            \in \BBR^{ 1 \times l_{\theta}},
     \end{align}
     and
     \begin{align}
        \pdt{\sigma(\SL(x, \theta)}{{x}}
            &=
                \pdt{\sigma}{{z}}
                \pdt{z}{\chi}
                \pdt{{\chi}}{x}
            =
                \sigma_z(z)
                \theta^\rmT 
                \SJ_{l_{x}}
            \in \BBR^{ 1 \times l_{x}}.
    \end{align}
\end{fact}

\begin{fact} 
    Let $z \isdef \SL(x, \Theta e) = \chi^\rmT \Theta e,$
    where $\Theta \in \BBR^{l_\theta \times \ell}.$
    Note that $e \in \BBR^{\ell}.$
    Then, 
    \begin{align}
        \pdt{\sigma(\SL(x, \Theta e_j)}{{\Theta}}
            &=
                \pdt{\sigma}{{z}}
                \pdt{z}{{\Theta}}
            =
                \sigma_z(z)
                e
                \chi^\rmT
            \in \BBR^{ \ell \times l_{\theta}}.
    \end{align}
\end{fact}
\vspace{1em}
\begin{fact}
    Let 
    \begin{align}
        N(x,\Theta) 
            =
                \matl
                    \sigma(\SL(x, \Theta e_1) \\
                    \vdots \\
                    \sigma(\SL(x, \Theta e_{\ell})
                \matr
                \in \BBR^{\ell},
    \end{align}
    where
    $\Theta = \matl \theta_1 & \cdots & \theta_\ell \matr \in \BBR^{l_x+1 \times \ell}.$
    Then, for $j=\{1, 2, \ldots, \ell\},$
    \begin{align}
        \pdt{N}{\theta_{j}}
            % =
            %     \pdt{}{\Theta_i}
            %     \matl
            %         \sigma(\SL(x, \Theta_i e_1) \\
            %         \vdots \\
            %         \sigma(\SL(x, \Theta_i e_{\ell_i})
            %     \matr
            &=
                \matl
                    \vdots \\
                    \pdt{\sigma(\SL(x, \theta_{j})}{{\theta_{j}}} \\
                    \vdots \\
                    % \pdt{\sigma(\SL(x_i, \theta_{i \ell_i} )}{{\theta_{ij}}}
                \matr
            % =
            %     \matl
            %         \vdots \\
            %         \sigma(z_{ij})
            %         \left( 1 - \sigma(z_{ij}) \right)
            %         x_i^\rmT \\
            %         \vdots
            %     \matr
            =
                \sigma_z(z_{j}) 
                % \left( 1 - \sigma(z_{ij}) \right) 
                e_j \chi^\rmT
                \in \BBR^{\ell \times l_{\theta}}
                ,
        \\
        \pdt{N}{x}
            &=
                \matl
                    \vdots \\
                    \pdt{\sigma(\SL(x, \theta_j)}{{x}} \\
                    \vdots
                \matr
            =
                \matl
                    \vdots \\
                    \sigma_z(z_{j})
                    % e_j^\rmT \Theta ^\rmT 
                    \theta_j^\rmT
                    \SJ_{l_x} \\
                    \vdots
                \matr
            \nn \\ &
            =
                \Lambda(x, \Theta) \Theta^\rmT \SJ_{l_x}
                \in \BBR^{\ell \times l_{x}  }
                ,
    \end{align}
    where 
    \begin{align}
        \Lambda(x, \Theta)
            =
                {\rm diag \ }
                (
                    \sigma_z(\SL(x, \theta_1)),
                    \ldots,
                    \sigma_z(\SL(x, \theta_\ell))
                )
            \in \BBR^{\ell \times \ell}.
        \label{eq:Lambda_def}
    \end{align}
\end{fact}
% \vspace{2em}

\begin{fact}
    \textbf{Backpropagation.}
    \label{fact:dy/dtheta}
    % Let $\theta_{ij} \in \BBR^{l_{\theta_i}} $ denote the $j$th column of $\Theta_i.$
    % Note that $\theta_{ij} =\Theta_i e_j,$ where $j \in \{1,2,\ldots, \ell_i\}.$
    Let $z_{ij} = \SL(x_i, \Theta_i e_j).$
    Then, for $i<n+1,$
    \begin{align}
        \pdt{y}{\theta_{ij}}
            &=
                \Theta_{n+1}^\rmT
                \prod_{q=n}^{i+1}
                \Lambda(x_{q}, \Theta_{q}) \Theta_{q}^\rmT \SJ_{l_{x_q}}
                \sigma_z(z_{ij}) e_j x_i^\rmT,
    \end{align}
    where $e_j$ is the $j$th column of the $\ell_i\times \ell_i$ identity matrix. 
    Furthermore, for $i=n+1,$
    \begin{align}
        \pdt{y}{\theta_{n+1,j}}
            &=
                \pdt{\Theta_{n+1}^\rmT x_{n+1}}{\theta_{n+1,j}}
            % =
            %     \pdt{}{\theta_{n+1,j}}
            %     \matl
            %         \vdots \\
            %         \theta_{n+1,j}^\rmT x_{n+1} \\
            %         \vdots
            %     \matr
            % \nn \\
            % &=
            %     \matl
            %         \vdots \\
            %         x_{n+1}^\rmT  \\
            %         \vdots \\
            %     \matr
            =
                e_j x_{n+1}^\rmT.
    \end{align}
    
\end{fact}

\begin{proof}
    Note that, for $i<n+1,$
    \begin{align}
        \pdt{y}{\theta_{ij}}
            &=
                % \pdt{\Theta_{n+1}^\rmT x_{n+1}}{\theta_{ij}}
                \pdt{S(x_{n+1})}{\theta_{ij}}
                % \pdt{N_n}{x_n}
            \nn \\
            &=
                \pdt{S(x_{n+1})}{x_{n+1}}
                \pdt{x_{n+1}}{\theta_{ij}}
            \nn \\
            &=
                \Theta_{n+1}^\rmT
                \pdt{N_n(x_{n},\Theta_n)}{\theta_{ij}}
            \nn \\
            &=
                \Theta_{n+1}^\rmT
                \pdt{N_n(x_{n},\Theta_n)}{x_n}
                \pdt{x_n}{\theta_{ij}}
            \nn \\
            &=
                \Theta_{n+1}^\rmT
                \Lambda(x_n, \Theta_n) \Theta_n^\rmT \SJ_{l_{x_n}}
                \pdt{N_{n-1}(x_{n-1},\Theta_{n-1})}{\theta_{ij}}
            \nn \\
            &=
                \Theta_{n+1}^\rmT
                \Lambda(x_n, \Theta_n) \Theta_n^\rmT \SJ_{l_{x_n}}
                \Lambda(x_{n-1}, \Theta_{n-1}) 
                \nn \\ &\quad 
                \cdot \Theta_{n-1}^\rmT \SJ_{l_{x_{n-1}}}
                \pdt{N_{n-2}(x_{n-2},\Theta_{n-2})}{\theta_{ij}}
            \nn \\
            &=
                \Theta_{n+1}^\rmT
                \prod_{q=n}^{i+1}
                \Lambda(x_{q}, \Theta_{q}) \Theta_{q}^\rmT \SJ_{l_{x_q}}
                \pdt{N_{i}(x_{i},\Theta_{i})}{\theta_{ij}}
            \nn \\
            &=
                \Theta_{n+1}^\rmT
                \prod_{q=n}^{i+1}
                \Lambda(x_{q}, \Theta_{q}) \Theta_{q}^\rmT \SJ_{l_{x_q}}
                \sigma_z(z_{ij}) e_j x_i^\rmT. \nn 
    \end{align}
\end{proof}

\end{document}

%% file: PackagesCommands.tex
\usepackage{amsmath} % assumes amsmath package installed
\usepackage{amssymb}  % assumes amsmath package installed
\usepackage{amsfonts}
\newtheorem{definition}{Definition}[section]
\newtheorem{exmp}{Example}[section]
\newtheorem{fact}{Fact}[section]
\usepackage{float} 
\usepackage[skip=1pt,font=footnotesize]{caption}
\usepackage[margin = 0.75in]{geometry}

\usepackage{subcaption}

\usepackage{etaremune}

\usepackage{tikz}
\usetikzlibrary{shapes,arrows,fit,calc,positioning}
\tikzstyle{bigblock} = [draw, fill=blue!20, rectangle, 
    minimum height=6em, minimum width=8em]
\tikzstyle{medblock} = [draw, fill=blue!20, rectangle, 
    minimum height=4em, minimum width=4em]    
\tikzstyle{mux} = [draw, fill=black!20, rectangle, 
    minimum height=5em, minimum width=0.1em]    
\tikzstyle{smallblock} = [draw, fill=blue!20, rectangle, 
    minimum height=2em, minimum width=3em]
    
\tikzstyle{data_block} = [draw, fill=green!20, rectangle, 
    minimum height=2em, minimum width=3em]
\tikzstyle{ops_block} = [draw, fill=blue!20, rectangle, 
    minimum height=2em, minimum width=3em]    
\tikzstyle{est_block} = [draw, fill=red!20, rectangle, 
    minimum height=2em, minimum width=3em]    
    
\tikzstyle{sum} = [draw, fill=blue!20, circle, node distance=1cm,minimum height=0.5cm]
\tikzstyle{signal} = [coordinate]
\tikzstyle{pinstyle} = [pin edge={to-,thin,black}]
\tikzstyle{block} = [draw, fill=blue!20, rectangle, 
    minimum height=3em, minimum width=9em]
\tikzstyle{blockS} = [draw, fill=blue!20, rectangle, 
    minimum height=3em, minimum width=4em]    
\tikzstyle{input} = [coordinate]
\tikzstyle{output} = [coordinate]
\usetikzlibrary{matrix}

\usetikzlibrary{positioning}
\usetikzlibrary{math}

\usepackage{hyperref}
\usepackage{xcolor}
\hypersetup{
    colorlinks,
    linkcolor={blue!100!black},
    citecolor={blue!50!black},
    urlcolor={blue!80!black}
}

\newcommand{\bc}{\begin{center}}
\newcommand{\ec}{\end{center}}
\newcommand{\benum}{\begin{enumerate}}
\newcommand{\eenum}{\end{enumerate}}
\newcommand{\nn}{\nonumber}
\newcommand{\matl}{\left[ \begin{array}}
\newcommand{\matr}{\end{array} \right]}
\newcommand{\matls}{\left[ \begin{smallmatrix}}
\newcommand{\matrs}{\end{smallmatrix} \right]}
\newcommand{\isdef}{\stackrel{\triangle}{=}}

\newcommand{\rmT}{{\rm T}}

\newcommand{\rme}{{\rm e}}

\newcommand{\BBR}{{\mathbb R}}

\newcommand{\SJ}{{\mathcal J}}

\newcommand{\SL}{{\mathcal L}}

\newcommand{\SN}{{\mathcal N}}

\renewcommand{\matl}{\begin{bmatrix}}
\renewcommand{\matr}{\end{bmatrix} }

\newcommand{\pdt}[2]{\ensuremath{\dfrac{\partial #1}{\partial #2}}}